\documentclass[prl,twocolumn,superscriptaddress,longbibliography]{revtex4-2}
\usepackage{amsmath}
\usepackage{amssymb}
\usepackage{amsthm}
\usepackage{amsfonts}
\usepackage{listings}
\usepackage{enumerate}
\usepackage{latexsym}
\usepackage{psfrag}
\usepackage{bm} 
\usepackage[all]{xy}
\usepackage{subfigure}
\usepackage{color}
\usepackage{mathtools}
\usepackage{array}
\usepackage{placeins}
\usepackage{hyperref}
\usepackage[export]{adjustbox}

\usepackage{physics}
\usepackage{tikz}
\usepackage{adjustbox}
\usetikzlibrary{calc}
\usetikzlibrary{decorations.pathreplacing}

\definecolor{OliveGreen}{cmyk}{0.64, 0, 0.95, 0.40}

\begin{document}

\title{1-Form Symmetric Projected Entangled-Pair States}

\author{Yi Tan}
\affiliation{Department of Physics, Southern University of Science and Technology, Shenzhen 518055, China}

\author{Ji-Yao Chen}
\email{chenjiy3@mail.sysu.edu.cn}
\affiliation{Center for Neutron Science and Technology, Guangdong Provincial Key Laboratory of Magnetoelectric Physics and Devices, School of Physics, Sun Yat-sen University, Guangzhou 510275, China}

\author{Didier Poilblanc}
% \email{didier.poilblanc@irsamc.ups-tlse.fr}
\affiliation{Laboratoire de Physique Th\'eorique, C.N.R.S. and Universit\'e de Toulouse, 31062 Toulouse, France}

\author{Fei Ye}
\affiliation{Department of Physics, Southern University of Science and Technology, Shenzhen 518055, China} 

\author{Jia-Wei Mei}
\email{meijw@sustech.edu.cn}
\affiliation{Department of Physics, Southern University of Science and Technology, Shenzhen 518055, China} 

\date{\today}
\begin{abstract}
The 1-form symmetry, manifesting as loop-like symmetries, has gained prominence in the study of quantum phases, deepening our understanding of symmetry. However, the role of 1-form symmetries in Projected Entangled-Pair States (PEPS), two-dimensional tensor network states, remains largely underexplored. We present a novel framework for understanding 1-form symmetries within tensor networks, specifically focusing on the derivation of algebraic relations for symmetry matrices on the PEPS virtual legs. Our results reveal that 1-form symmetries impose stringent constraints on tensor network representations, leading to distinct anomalous braiding phases carried by symmetry matrices. We demonstrate how these symmetries influence the ground state and tangent space in PEPS, providing new insights into their physical implications for enhancing ground state optimization efficiency and characterizing the 1-form symmetry structure in excited states. 
\end{abstract}
\maketitle

\emph{Introduction --}
Quantum many-body systems are challenging to study phases of matter due to complex interactions and entanglements\cite{Wen2004,Amico2008,Eisert2010,Cirac2021}. Projected Entangled-Pair States (PEPS) offer a powerful two-dimensional tensor-network framework for efficiently representing these states\cite{Verstraete2004,Cirac2021}. By encoding symmetries, tensor-network representations classify phases of matter, particularly symmetry-protected topological phases\cite{Chen2011,Chen2012,Pollmann2010,Pollmann2012}. Additionally, virtual symmetries (gauge symmetries) are crucial in characterizing topological phases, defining properties like topological entanglement entropy, modular matrices, and anyonic excitations\cite{Gu2008,Schuch2010,Cirac2021,Mei2017}.

Recently, higher-form symmetries, especially 1-form symmetries, have gained prominence in the study of quantum phases\cite{Nussinov2009,Gaiotto2015,Wen2019,McGreevy_2023,Choi2022,Qi2021,Cordova2022,Ellison2023}. These symmetries, which manifest as loop-like symmetries through string operators, extend global (0-form) symmetries  to higher-dimensional objects, providing a deeper understanding of symmetry in quantum systems\cite{Gomes2023,Schafernameki2023}. The study of higher-form symmetries has led to significant insights in quantum computing\cite{Ellison2023}, topological phases of matter\cite{Wen2019,McGreevy_2023,Qi2021}, and quantum gravity\cite{Harlow2021}. They are crucial for understanding the dynamics of quantum field theories\cite{Cordova2022}, including anomaly constraints and the behavior of exotic phases of matter\cite{Ma2023,Liu2024}.

Kitaev's spin-1/2 model on the honeycomb lattice\cite{Kitaev2006} lays the groundwork for various Abelian and non-Abelian topologically ordered states and inspires potential material realizations~\cite{Jackeli2009,Chaloupka2010,Takagi2019}. This model and its higher-spin extensions\cite{Baskaran2008} exhibit a lattice version of \(\mathbb{Z}_2^{[1]}\) 1-form symmetry\cite{Baskaran2008,Ma2023,Liu2024}, which can be generalized to \(\mathbb{Z}_N^{[1]}\) 1-form symmetry\cite{Ellison2023,Liu2024} in the \(\mathbb{Z}_N\) Kitaev model\cite{Barkeshli2015,Chen2024}. The anomalous 1-form symmetry ensures fractionalized excitations in the ground state of the Kitaev model even without exact solutions\cite{Baskaran2008,Ellison2023,Ma2023,Liu2024}. Despite these advancements, the role of 1-form symmetries in PEPS remains largely unexplored. This work formally devises the framework of 1-form symmetric PEPS. We derive algebraic relations for symmetry matrices acting on the virtual legs of PEPS along loops, demonstrating how these matrices capture fractionalized anyonic excitations enforced by anomalous 1-form symmetries. We present concrete solutions to these algebraic equations for the $\mathbb{Z}_N$ Kitaev models on the honeycomb lattice and develop a 1-form symmetric tangent space. By demonstrating enhanced ground state optimization efficiency and characterizing the 1-form symmetry structure in excited states, we showcase the crucial role of 1-form symmetries in both ground and excited states of PEPS.

\emph{Generic 1-form symmetric PEPS representation --}
On a two-dimensional manifold that admits a decomposition into a lattice composed of sites, edges, and plaquettes
\[
\includegraphics[width=0.48\columnwidth,valign=c]{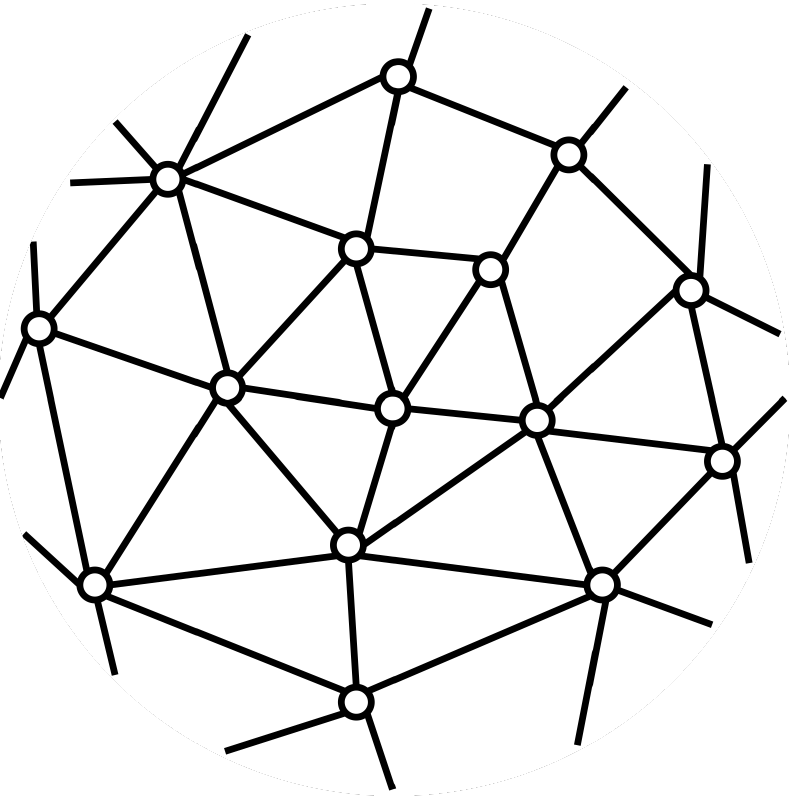},
\]
we can define tensor-network states as
\[|\Psi\rangle = \sum_{s_1,\cdots,s_N}{\rm tTr} A^{(i)s_i}_{abcde\cdots}|s_1\cdots s_i\cdots s_N\rangle,\]
where the tensor trace \(\mathrm{tTr}\) denotes the contraction of virtual indices over the site-dependent local tensor \(A^{(i)s_i}_{abcde\cdots}=\includegraphics[width=0.2\columnwidth,valign=c]{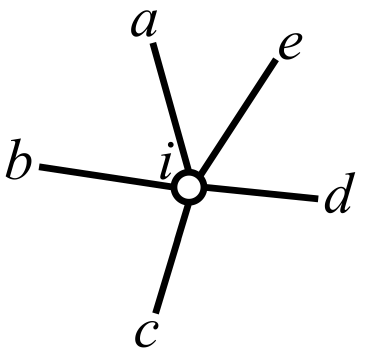}\), which has the physical index \( s_i \) on the \( i \)-th site, and virtual indices \( a, b, c, d, e, \ldots \) corresponding to the edges of the lattice.

We consider the \(\mathbb{Z}_N^{[1]}\) 1-form symmetry which manifests as loop-like symmetries through
loop operators on the manifold. For an arbitrary loop \(\gamma\) among the loops on the manifold
  \[
\includegraphics[width=0.48\columnwidth,valign=c]{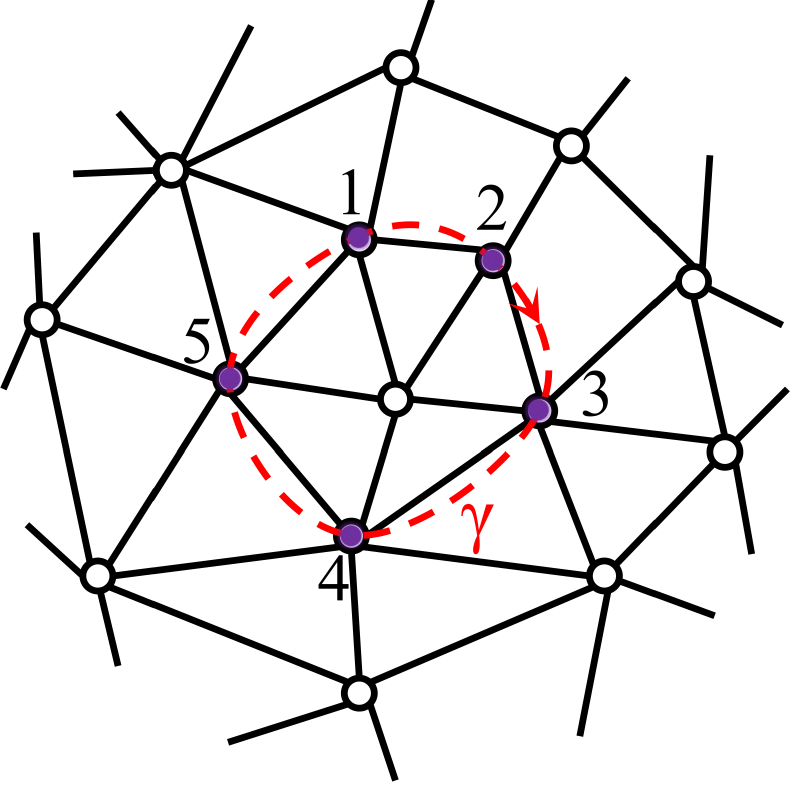},
\]
the loop operator is defined as
\begin{equation}\label{eq:gamma}
W_\gamma = U^{(1)}_\gamma U^{(2)}_\gamma U^{(3)}_\gamma U^{(4)}_\gamma U^{(5)}_\gamma,
\end{equation}
where each \( U^{(i)}_\gamma \) represents the physical operator at the \( i \)-th site along the loop \(\gamma\). 

We examine the relevant local tensors along the \(\gamma\)-loop, and interpret \(\includegraphics[width=0.28\columnwidth,valign=c]{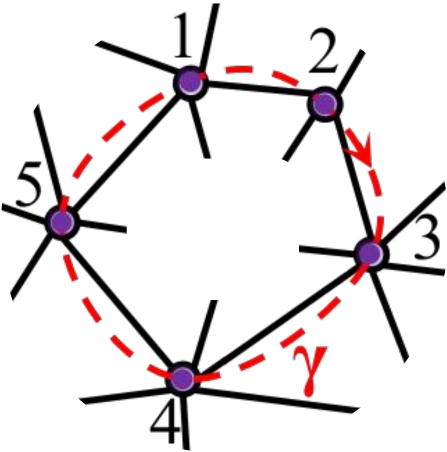}\) as a matrix product state (MPS) as the legs (both physical and virtual) that are not along the loop are treated as the ``physical'' ones. Similar to the string operators in the symmetric MPS\cite{Cirac2021}, the loop symmetry operators can be projectively represented with the projective phase $\theta$
\begin{eqnarray}\label{eq:1form}
U_{\gamma,ss'}^{(i)} A^{(i)s'}_{abcd} = e^{i \theta} A^{(i)s}_{ab'cd'} V^{(il)}_{bb'} V^{(ir)}_{dd'},
\end{eqnarray}
which has the graph representation
\[\includegraphics[width=0.48\columnwidth,valign=c]{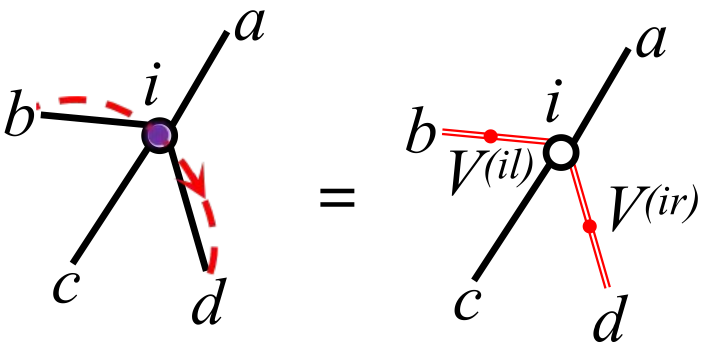}.\]
Here, the matrices \(V^{(il)}_{cc'}\) and \(V^{(ir)}_{ee'}\) represent the virtual symmetry operations on the left and right virtual legs of the \(i\)-th tensor along the loop \(\gamma\), respectively. The symmetry matrices on the \emph{partial} virtual legs is also explored in the context of stabilizer PEPS\cite{Herringer2023}, including the toric code PEPS, without explicitly referring to 1-form symmetry, although it is implicitly encoded as the subsystem symmetry\cite{Stephen2019}.

Thus, the 1-form symmetry operator \(W_\gamma\) acting on the physical indices can be effectively translated into the symmetry matrices on the virtual legs of the loop up to a projective phase
\[\includegraphics[width=0.48\columnwidth,valign=c]{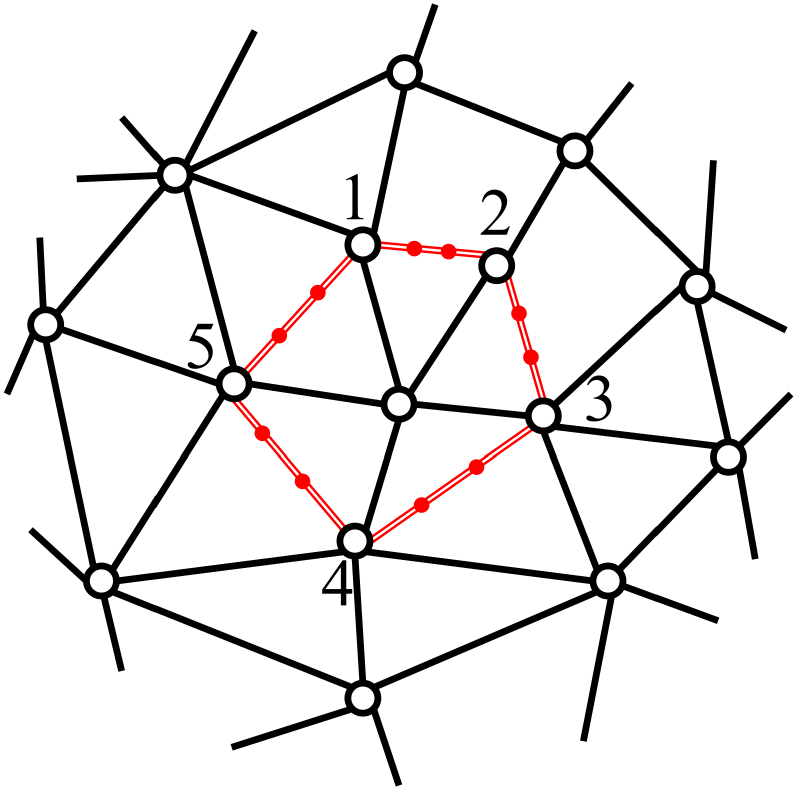}.\]
A gauge consistency condition \(\includegraphics[width=0.2\columnwidth,valign=c]{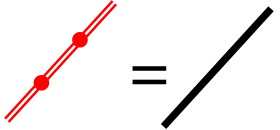}\) is required for the virtual symmetry matrices
\begin{eqnarray}\label{eq:gauge}
    (V^t)^{(ir)} V^{(i+1 \, l)}=\mathbf{1}.
\end{eqnarray}
Here the superscript \(t\) denotes the transpose of the matrix and \(\mathbf{1}\) is the identity matrix.
With a proper choice of the projective phase, we construct the zero-flux state with $W_\gamma|\psi\rangle=|\psi\rangle$. 
To construct a faithful representation of \( W_{\gamma} \), we introduce a ``charge'' (flux) operator \({g}_\gamma\) that satisfies the commutation relation
\(
{g}_\gamma W_\gamma = w W_\gamma {g}_\gamma
\)
with \( w^N = 1 \) to increases the flux of \( W_\gamma \) enclosed by the loop \(\gamma\). Using this charge operator, we can construct eigenstates with different flux values
\[
W_\gamma ({g}_\gamma)^n |\psi\rangle = w^n ({g}_\gamma)^n |\psi\rangle.
\]

In the context of PEPS, the flux operator can be represented by multiplying a matrix \(\mathfrak{g}_{\rm flux}\) on any bond along the \(\gamma\)-loop, e.g. the third bond shared by the third and fourth local tensors, as depicted below
\[
\includegraphics[width=0.65\columnwidth,valign=c]{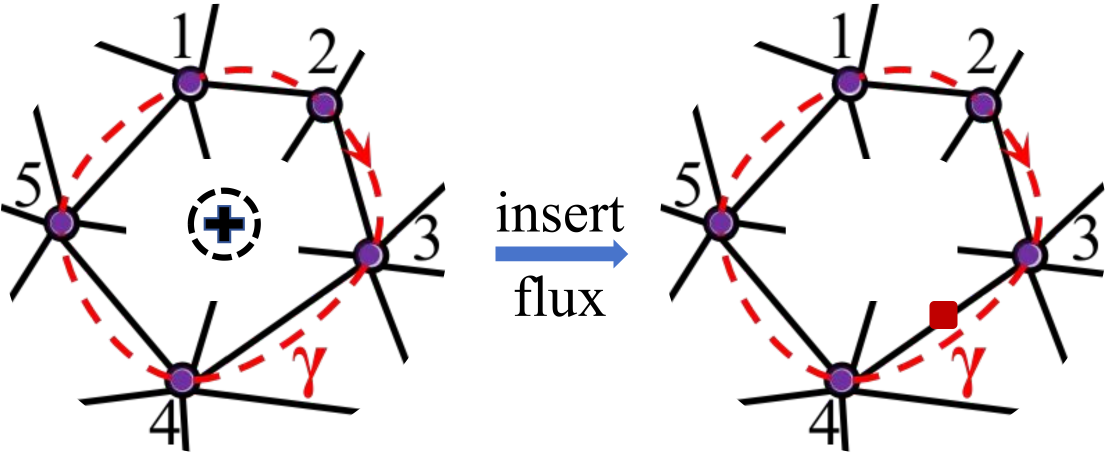}.
\]
We can apply the flux operator \(\mathfrak{g}_{\rm flux}\) to either the third or the fourth tensor, requiring the consistency condition
\begin{eqnarray}
\label{eq:flux_OP}
    \mathfrak{g}^{(i)t}_{\rm flux} \mathfrak{g}^{(i)}_{\rm flux} = \mathbf{1}.
\end{eqnarray}
To ensure that the flux state is an eigenstate, the flux condition \(\includegraphics[width=0.18\columnwidth,valign=c]{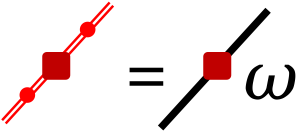}\) must be satisfied,
which translates to
\begin{eqnarray}\label{eq:flux}
(V^t)^{(ir)} \mathfrak{g}^{(i)}_{{\rm flux}} V^{(i+1 \, l)} = w \mathfrak{g}^{(i)}_{{\rm flux}}.
\end{eqnarray}

While \(W_\gamma\) is loop dependent, the virtual symmetry matrices on the legs of tensors are not. Therefore, it is necessary to check the consistency conditions for loops (e.g., \(\gamma\) and \(\delta\)) sharing common edges, such as the edge (3-4) in the following diagram
\[
\includegraphics[width=0.48\columnwidth,valign=c]{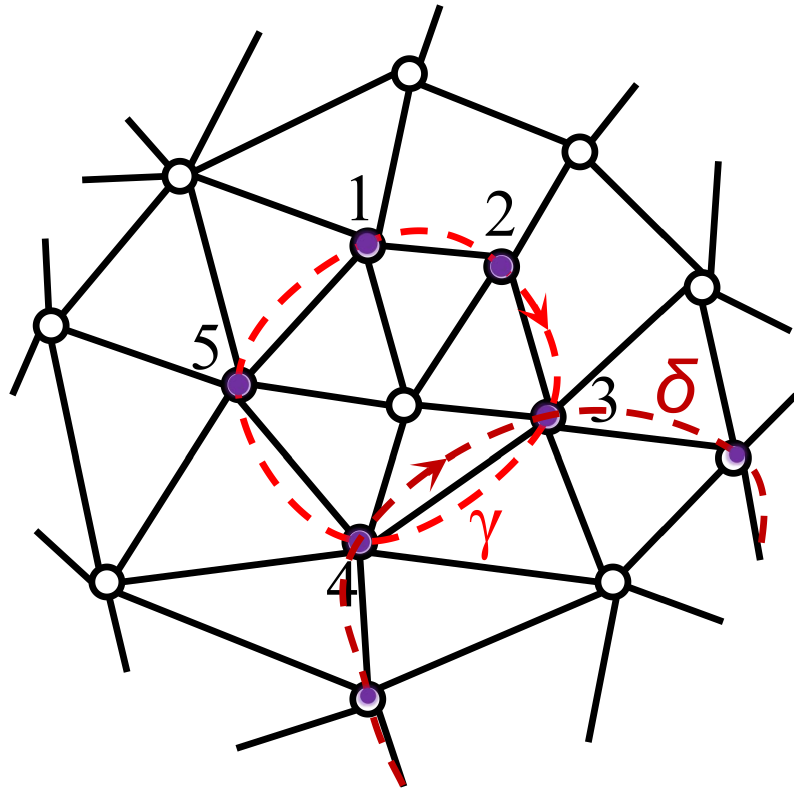}.
\]
The commutation relation  \([W_\gamma, W_\delta] = 0\) requires
\[
U_{\gamma}^{(3)} U_{\gamma}^{(4)} U_{\delta}^{(3)} U_{\delta}^{(4)} = U_{\delta}^{(3)} U_{\delta}^{(4)} U_{\gamma}^{(3)} U_{\gamma}^{(4)},
\]
leading to the phase factor
\[
\omega_U =(U_\gamma^{(3)} U_\delta^{(3)})^\dagger U_\delta^{(3)} U_\gamma^{(3)} = U_\gamma^{(4)} U_\delta^{(4)} (U_\delta^{(4)} U_\gamma^{(4)})^\dagger,
\]
which can be used to classify the \(\mathbb{Z}_N\) 1-form symmetry protected topological states\cite{Tsui2020,Inamura2024}. 
Therefore, the virtual symmetry operators also have the non-commutative \(U(1)\) phase
\begin{eqnarray}\label{eq:phase}
V^{(i)}_{\gamma} V^{(i)}_{\delta} = \omega_U V^{(i)}_{\delta} V^{(i)}_{\gamma},
\end{eqnarray}
representing the braiding phase of excitations.

The loop-like operators can be arbitrarily deformed under 1-form symmetry. For the nearby loops around the joint \(0\)th-point
\[
\includegraphics[width=0.36\columnwidth,valign=c]{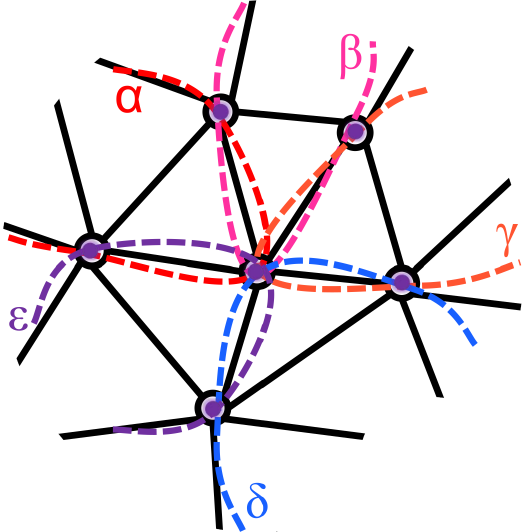}, 
\]
the deformability requires
\[
U_\alpha^{(0)}U_\beta^{(0)}U_\gamma^{(0)}U_\delta^{(0)}U_\epsilon^{(0)} = 1
\]
up to a U(1) phase, which results in a gauge symmetry acting only on all the virtual legs of the local tensor
\begin{equation}\label{eq:gauge_symm}
\includegraphics[width=0.48\columnwidth,valign=c]{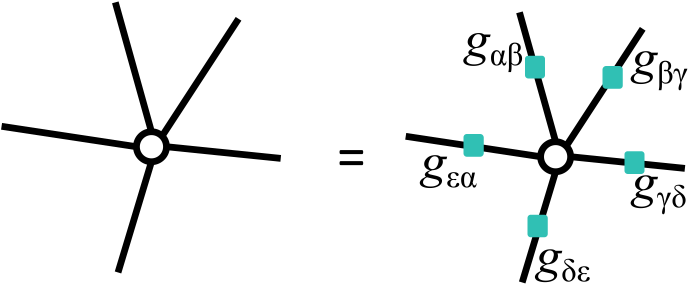},
\end{equation}
with e.g. \(g_{\alpha\beta} = V^{(0)}_\beta V^{(0)}_\alpha\). Therefore, the deformability of the 1-form symmetry implies a gauge symmetry when \(g_{\alpha\beta}\) is not the identity matrix.

\emph{1-form symmetric PEPS for Kitaev-type model --}
The 1-form symmetry loop operators in Eq. (\ref{eq:gamma}) acting on physical degrees of freedom can be represented by symmetry matrices in Eq. (\ref{eq:1form}) operating on the virtual legs. To construct the 1-form symmetric PEPS, we need determine the virtual symmetry matrices by solving the algebraic relations outlined in Eqs. (\ref{eq:gauge}), (\ref{eq:flux_OP}), (\ref{eq:flux}), and (\ref{eq:phase}). 
We will present the concrete solutions for the necessary algebraic equations for the symmetry matrices of the \(\mathbb{Z}_N^{[1]}\) 1-form symmetry on the honeycomb lattice for the \(\mathbb{Z}_N\) Kitaev model.

For the translational PEPS on the honeycomb lattice
\(|\Psi(A)\rangle = \sum_{s_1,\cdots,s_N}{\rm tTr} A_{abc}^{s_i}|s_1\cdots s_i\cdots s_N\rangle=\includegraphics[width=0.45\columnwidth,valign=c]{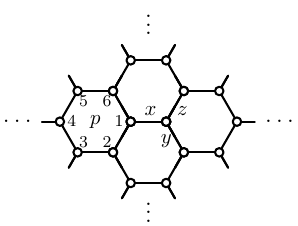}\),
the local tensor \(A_{abc}^{s}\) comprises one physical index (\(s\)) and three virtual indices (\(a\), \(b\), and \(c\)) along the nearest-neighbor \(x\), \(y\), and \(z\) bonds.
Encircling a single hexagon, the 1-form symmetry loop operator\cite{Kitaev2006a,Baskaran2008,Barkeshli2015,Ellison2023}  is 
\begin{equation}\label{eq:Wp}
W_p = U^{(1)}_x U^{(2)}_y U^{(3)}_z U^{(4)}_x U^{(5)}_y U^{(6)}_z.     
\end{equation}
The 1-form symmetry Eq.(\ref{eq:1form}) in PEPS requires
\begin{eqnarray}\label{eq:1form_hc}
    U_{x,ss'} A_{ijk}^{s'} &=& e^{i\theta} Y_{x,jj'}Z_{x,kk'}A_{ij'k'}^s \nonumber \\
    U_{y,ss'} A_{ijk}^{s'} &=& e^{i\theta} Z_{y,kk'}X_{y,ii'}A_{i'jk'}^s \nonumber \\
    U_{z,ss'} A_{ijk}^{s'} &=& e^{i\theta} X_{z,ii'}Y_{z,jj'}A_{i'j'k}^s,
\end{eqnarray}
to ensure that the loop symmetries are accurately represented. Here we assume the virtual symmetry matrices for local tensors on different sublattices are the same.
The gauge condition in Eq.(\ref{eq:gauge}) leads to 
\begin{equation}\label{eq:unitary}
Z_y=Z_x^*, X_z=X_y^*, Y_x=Y_z^*,    
\end{equation}
as we require unitarity. 

For the \(\mathbb{Z}_N^{[1]}\) 1-form symmetry, the \(N\)-dimensional virtual symmetry matrices are chosen from Sylvester's generalized \(N\)-dimensional Pauli matrices with \(\sigma_x = \sum_{\alpha}|\alpha+1\rangle\langle\alpha|\), \(\sigma_z = \sum_{\alpha}\omega^\alpha|\alpha\rangle\langle\alpha|\) with \(\omega = e^{2\pi i/N}\), and the \(y\)-Pauli operator as \(\sigma_y = \sigma_x^\dag \sigma_z^\dag\) for odd \(N\), and \(\sigma_y = \sqrt{\omega}\sigma_x^\dag \sigma_z^\dag\) for even \(N\)\cite{Kibler2008,Ellison2023}. From the flux conditions in  Eqs.(\ref{eq:flux_OP}) and (\ref{eq:flux}), we specify 
\begin{eqnarray}\label{eq:w}
   \mathfrak{g}_{\rm flux} = \sigma_x, Z_x^\dag\sigma_xZ_x=X_y^\dag\sigma_xX_y=Y_z^\dag\sigma_xY_z=\omega^*\sigma_x,
\end{eqnarray}
and thus \(X_y\), \(Y_z\), and \(Z_x\) can be expressed in terms of \(\sigma_x^a \sigma_z\)  with integer \(a\)\cite{Kibler2008}, e.g. \(X_y = \sum_a c_a \sigma_x^a \sigma_z\). The commutation relation in Eq.(\ref{eq:phase}) leads to
\begin{equation}\label{eq:wU}
    Z_x Z_x^*=\omega_U Z_x^* Z_x,X_yX_y^*=\omega_U X_y^*X_y, Y_z Y_z^*=\omega_U Y_z^*Y_z, 
\end{equation} 
implying
\[
\sum_{ab} c_a c_b^* \omega^b \sigma_x^{a+b} = \omega_U \sum_{ab} c_a^* c_b \omega^{-b} \sigma_x^{a+b}.
\]

For \(\mathbb{Z}_N^{[1]}\) 1-form symmetry, the braiding phase \(\omega_U\) can be expressed as \(\omega_U = \omega^{-u}\), where \(u\) is an integer. An explicit solution is
\begin{equation}\label{eq:sol}
    X_y = Y_z = Z_x = \sigma_x^{(N-u)/2} \sigma_z,
\end{equation}
when \((N-u)/2\) is an integer. For non-integer values of \((N-u)/2\), our attempt to find a general explicit solution is unsuccessful. It is important to note that we do not assume uniqueness and existence of the solution, which remains open and warrants further investigation.

For the integer spin-\(S\) Kitaev model\cite{Baskaran2008,Ma2023,Liu2024} where \(N=2\) and \(u=2\), \((N-u)/2=0\), the corresponding solution is
\begin{equation}\label{eq:spin_one}
    X_y = Y_z = Z_x = X_z = Y_x = Z_y = \sigma_z.
\end{equation}
For the \(\mathbb{Z}_3\) Kitaev model\cite{Barkeshli2015,Ellison2023,Liu2024,Chen2024} where \(N=3\) and \(u=1\), \((N-u)/2=1\), the corresponding solution is
\begin{equation}\label{eq:z3}
    X_y = Y_z = Z_x = \sigma_y, \quad X_z = Y_x = Z_y = \sigma_y^*.
\end{equation}
For the half-integer spin-\(S\) Kitaev model\cite{Kitaev2006a,Baskaran2008,Ma2023,Liu2024} where \(N=2\) and \(u=1\), \((N-u)/2=1/2\) is not an integer, making Eq. (\ref{eq:sol}) inapplicable. However, we have identified the following solution
\begin{equation}\label{eq:spin_half}
    X_y = Y_z = Z_x = \frac{\sigma_y + \sigma_z}{\sqrt{2}},\quad X_z = Y_x = Z_y = \frac{\sigma_y^* + \sigma_z}{\sqrt{2}}.
\end{equation}
Notably, the symmetry matrices in Eqs.(\ref{eq:spin_half}) and (\ref{eq:spin_one}) are identical to those for the symmetric variational ground PEPS for the spin-\(S\) model with a rotated basis\cite{Lee2019,Lee2020,Lee2021,Tan2024}.

The braiding phase \(\omega_U\) is a critical indicator of the 1-form symmetry anomaly\cite{Tsui2020,Inamura2024,Ma2023, Liu2024,Ellison2023, Chen2024}. In the integer spin-\(S\) Kitaev model, the braiding phase is trivial as seen from the commutative nature of the solution in Eq. (\ref{eq:spin_one}), such as \(X_y\) and \(X_z\), which implies a bosonic anyon related to the 1-form symmetric operator. Conversely, for half-integer spins, the solution in Eq. (\ref{eq:spin_half}) involves anti-commutative matrices like \(X_y\) and \(X_z\), indicating a non-trivial (fermionic) braiding phase for the anyons. This distinct even-odd effect for half-integer spins is discussed in detail in Refs. \cite{Ma2023, Liu2024}. 
The \(\mathbb{Z}_N^{[1]}\) 1-form symmetry for \(N \geq 3\) is further examined in Refs. \cite{Ellison2023, Chen2024}. The anomaly is explicitly reflected in the solution in Eq. (\ref{eq:z3}), where \(X_y\) and \(X_z\) exhibit a non-trivial commutation braiding phase.
The pure gauge symmetry matrices \(g_{\rm gauge} = X_y X_z = Y_x Y_z = Z_x Z_y\) according to Eq.(\ref{eq:gauge_symm}) are non-trivial for the solutions in Eqs. (\ref{eq:spin_half}) and (\ref{eq:z3}), indicating non-trivial topological order in the ground state, while trivial for solutions in Eq. (\ref{eq:spin_one}), further confirming the different 1-form symmetry anomaly behaviors.

\emph{1-form symmetric tangent space --}
The PEPS tangent space describes variations and excitations around a given PEPS configuration and captures the local structure and dynamics\cite{Vanderstraeten2015}. The set of uniform PEPS defines a manifold through the mapping between the set of local tensors \(A\) and the physical states in the Hilbert space \(|\Psi(A)\rangle\). For the PEPS on the honeycomb lattice, an overcomplete basis for a tangent vector is obtained as
\(|\Phi(B;A)\rangle = \sum_i |B_i\rangle\) with
\(
|B_i\rangle = \sum_{abca'b';s_{\alpha_i},s_{\beta_i}} B_{aba'b'}^{s_{\alpha_i},s_{\beta_i}} \frac{\partial}{\partial (A_{abc}^{s_{\alpha_i}} A_{a'b'c}^{s_{\beta_i}})} |\Psi(A)\rangle=\includegraphics[width=0.45\columnwidth,valign=c]{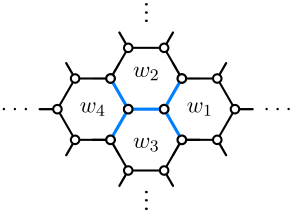}
\).
For the 1-form symmetric PEPS, the tangent vector can be symmetrized using the eigen-equations \(\hat{W}_p |B_i\rangle = w_p N_B |B_i\rangle\) for the surrounding four plaquette operators \(\hat{W}_{1,2,3,4}\), where $W_p$ is projection of plaquette operator onto variational space and \(N_B\) is the norm matrix of \(|B_i\rangle\)\cite{Tan2024}. We select \(|B_i\rangle\) with \(w_{1,2,3,4}\) the same as the ground state for the basis of the 1-form symmetric tangent space and expand the local tensor as \( A = \sum_k B_i^k x_k \), leading to the tangent-space gradient \( G_k = 2 \frac{\partial e(A, \bar{A})}{\partial \bar{x}_k} = \text{tTr}((B_i^k)^\dag g) \) with \( g = 2 \frac{\partial e(A, \bar{A})}{\partial \bar{A}} \) with the variational energy \(e\)\cite{Vanderstraeten2016}.

\begin{figure}[t]
    \centering
    \includegraphics[width=\columnwidth]{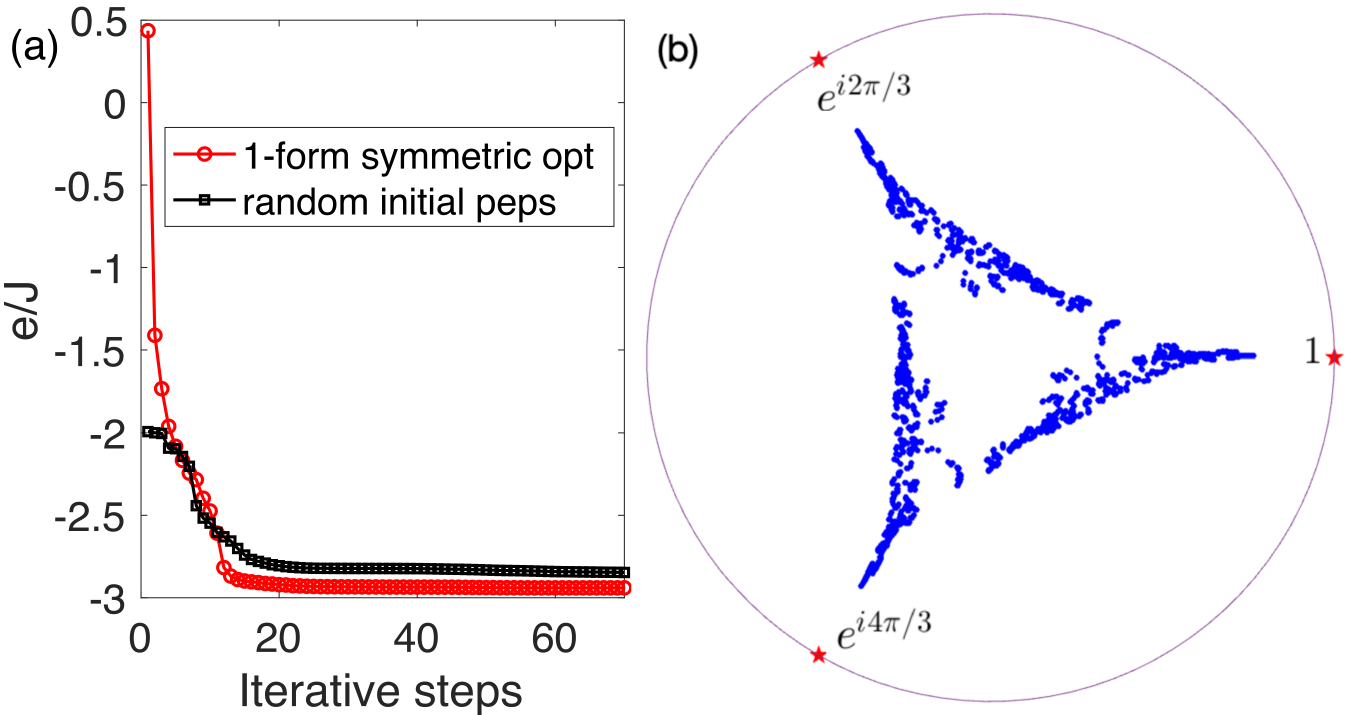}
    \caption{(a) Gradient-based optimization of the \(\mathbb{Z}_3\) Kitaev ground state. (b) Eigenvalue distribution of the loop operator \(W_p\) for the \(\mathbb{Z}_3\) model. The red stars represent the expected eigenvalues \(\omega = 1, e^{\pm i2\pi/3}\) for the 1-form symmetric PEPS. The other points denote the eigenvalues obtained from the optimized wave function starting from a randomly initialized PEPS.
    }
    \label{fig:optim}
\end{figure}

For the implications of 1-form symmetric PEPS, we performed a PEPS simulation of the ground and local excited states for the \(\mathbb{Z}_3\) Kitaev model \cite{Barkeshli2015, Chen2024} with 
\(
H = -J \sum_{\langle i,j\rangle \in \alpha} \sigma_i^\alpha \sigma_j^\alpha + {\rm H.C.}
\). Initially, a non-symmetric gradient-based optimization starting from a randomly initialized PEPS with bond dimension \(D=6\) fails to converge to a satisfactory variational state, yielding the eigenvalue distribution of the loop operator \(W_p=(\sigma_x^1\sigma_y^2\sigma_z^3\sigma_x^4\sigma_y^5\sigma_z^6)^\dag\) for all local excited states, as depicted in Fig.~\ref{fig:optim}. The results, far from the expected values of \(\omega = 1, e^{\pm i2\pi/3}\), indicate the presence of local minima, preventing the reliable acquisition of the 1-form symmetric ground state.

To overcome this, we utilize the 1-form symmetric tangent gradient to perform a symmetric optimization. We employ symmetry matrices derived from the kronecker product of the solution of Eq.~(\ref{eq:z3}) and a \(2 \times 2\) identity matrix to symmetrize the PEPS, resulting in an initial 1-form symmetric PEPS. As shown in Fig.~\ref{fig:optim}, the 1-form symmetric optimization significantly improved optimization efficiency and successfully converged to the eigenstate of the 1-form symmetry operator, achieving the anticipated eigenvalues \(\omega = 1, e^{\pm i2\pi/3}\) for all the local excited states. Our results align with previous DMRG simulations which identified the chiral ground state with \(w_p = 1\)\cite{Chen2024}, demonstrating the efficiency of the 1-form symmetric PEPS approach.

\emph{Conclusion --}
In this study, we have introduced a novel framework for understanding 1-form symmetries within tensor networks, with a specific focus on Projected Entangled-Pair States (PEPS). Through the derivation of algebraic relations for symmetry matrices on the PEPS virtual legs along loops associated with 1-form symmetry, we have demonstrated that 1-form symmetries impose stringent constraints on tensor network representations. This results in distinct anomalous braiding phases carried by the symmetry matrices, enabling 1-form symmetric PEPS to effectively capture the anomaly in the 1-form symmetry.

We have underscored the potential for enhancing ground state optimization efficiency and characterizing the 1-form symmetry structure in excited states. Although we demonstrate the implications of 1-form symmetry in the Kitaev honeycomb model, these findings can be straightforwardly applied to other systems, such as the 1-form symmetry in the Kitaev star lattice model\cite{Yao2007,Lee2020a}, where the same algebraic equations and solutions hold valid. Previous numerical studies of the Kitaev models have found significant entanglement spectrum degeneracy\cite{Shinjov2015,Lee2020a,Chen2024}, which is likely characterized by the 1-form symmetry for the boundary operators, as defined in the Wen plaquette model\cite{Wen2003,Ho2015}. The 1-form symmetry opens new avenues for the PEPS study of topological phases in strongly correlated systems.

\section{Acknowledgment}
This work is dedicated to the memory of T.~M. Rice, in gratitude for his invaluable teaching on the physics of strongly correlated systems. J.~W. Mei thanks Liujun Zou for useful discussions. 
J.-Y. C. would like to thank Meng Cheng for collaboration on related topics.
This work is supported by the National Key Research and Development Program of China (Grant No.~2021YFA1400400), Shenzhen Fundamental Research Program (Grant No.~JCYJ20220818100405013  and JCYJ20230807093204010). 
J.-Y. C. was supported by  National Natural Science Foundation of China (Grant No.~12304186), Open Research Fund Program of the State Key Laboratory of Low-Dimensional Quantum Physics (Project No.~KF202207), Innovation Program for Quantum Science and Technology 2021ZD0302100, Guangzhou Basic and Applied Basic Research Foundation (Grant No.~2024A04J4264), and Guangdong Basic and Applied Basic Research Foundation (Grant No.~2024A1515013065).
Part of the calculations reported were performed on resources provided by the Guangdong Provincial Key Laboratory of Magnetoelectric Physics and Devices, No.~2022B1212010008.
This work was also supported by the TNTOP ANR-18-CE30-0026-01 grant awarded by the French Research Council.

\bibliography{draft}

\end{document}